\documentclass{article}

\usepackage[preprint]{neurips_2026}
\usepackage[utf8]{inputenc}
\usepackage[T1]{fontenc}
\usepackage{amsmath,amssymb}
\usepackage{booktabs}
\usepackage{microtype}
\usepackage{hyperref}
\usepackage{url}
\usepackage{xcolor}
\usepackage{multicol}
\usepackage{enumitem}
\usepackage{tikz}
\usepackage{booktabs,tabularx,array}
\usetikzlibrary{arrows.meta,calc}
\usepackage{flafter}
\usepackage{placeins}
\definecolor{jevblue}{HTML}{305E91}
\definecolor{jevgreen}{HTML}{247D70}
\definecolor{jevpurple}{HTML}{79549B}
\definecolor{jevink}{HTML}{243E53}
\definecolor{jevgray}{HTML}{637380}
\hypersetup{colorlinks=true,citecolor=blue,linkcolor=blue,urlcolor=blue}

\AtBeginDocument{%
  \setlength{\abovedisplayskip}{6.5pt plus 2pt minus 2pt}%
  \setlength{\belowdisplayskip}{6.5pt plus 2pt minus 2pt}%
  \setlength{\abovedisplayshortskip}{3pt plus 1pt minus 1pt}%
  \setlength{\belowdisplayshortskip}{5pt plus 1pt minus 1pt}%
}

\title{Qwen-Audio-Jev:\\A Unified Interface for Audio Decisions}
\author{%
  \raisebox{-0.2\height}{\includegraphics[height=1.2em]{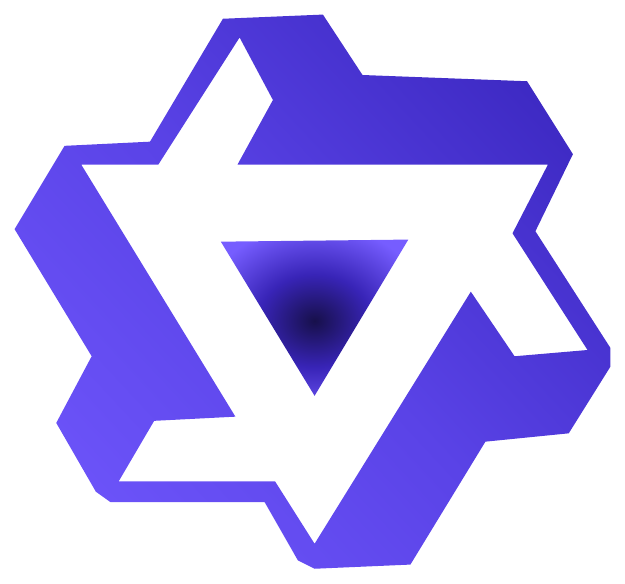}}\hspace{0.55em}Token Foundry, Alibaba Group\\[3pt]
  {\normalfont\normalsize\href{https://qwenaudio.github.io/qwen-audio-jev/}{Project Page}}%
}

\makeatletter
\renewcommand{\@noticestring}{%
  Technical report.\quad GitHub: \url{https://github.com/QwenAudio/Qwen-Audio-Jev}\\
  Hugging Face weights:
  \href{https://huggingface.co/FunAudioLLM/Qwen-Audio-3.1-Jev}{Qwen-Audio-3.1-Jev}\quad
  \href{https://huggingface.co/FunAudioLLM/Qwen-Audio-3.1-Jev-Tiny}{Qwen-Audio-3.1-Jev-Tiny}%
}
\makeatother

\begin{document}
\maketitle
\raggedbottom

\begin{abstract}
Large audio language models (LALMs) have made substantial progress in audio understanding. However, autoregressive text generation introduces computational overhead and latency in high-throughput applications that require frequent selection, judgment, and scoring. To address this limitation, we introduce Qwen-Audio-Jev in two variants: Qwen-Audio-3.1-Jev for high-accuracy audio decisions and Qwen-Audio-3.1-Jev-Tiny for resource-efficient complete-clip and streaming decisions. Both augment pretrained LALMs with a dedicated Jev decision head and unify selection, binary judgment, and scoring through three decision primitives: \textsc{Choice}, \textsc{Noul}, and \textsc{Score}. Given a natural-language question and answer specification followed by the audio and a decision token, Qwen-Audio-Jev directly outputs a task-conditioned probability distribution or a numerical score through typed decision readouts. Evaluation across 32 dataset--task configurations characterizes Qwen-Audio-3.1-Jev: \textsc{Choice} achieves 79.23\% accuracy on MMAU and 82.38\% on MMSU; \textsc{Noul} achieves 94.20\% accuracy on the VoxCeleb1-O speaker verification benchmark; and \textsc{Score} achieves a normalized mean absolute error of 0.1110 on NISQA overall quality assessment. We also implement streaming support in Qwen-Audio-3.1-Jev-Tiny, enabling successive decision readouts as audio arrives.

\end{abstract}

\section{Introduction}
\label{sec:introduction}
Audio carries linguistic content, speaker characteristics, vocal expression, acoustic events, and temporal structure. Applications draw on this information to identify sounds, verify speaker identity, and assess speech quality. Although these tasks differ in semantics, their outputs often take a bounded form: a category, a binary judgment, or a numerical value on a predefined scale. In high-throughput settings, a model must produce these decisions repeatedly and efficiently.

Large audio language models (LALMs) have substantially expanded the scope of audio understanding. Qwen2.5-Omni supports multimodal perception and streaming text and speech generation~\citep{xu2025qwen25omni}; Qwen3-Omni adopts a Thinker--Talker MoE architecture~\citep{xu2025qwen3omni}, and Qwen3.5-Omni extends this family with hybrid-attention MoE and dynamic text--speech alignment~\citep{qwen2026qwen35omni}. SALMONN integrates speech and general-audio encoders with a language model~\citep{tang2024salmonn}; LTU and GAMA investigate audio question answering and instruction tuning~\citep{ghosh2024gama,gong2024ltu}. Qwen-Audio-3.1-ASR-Flash-Next integrates speech recognition, speaker-diarized transcription, and general audio understanding~\citep{qwenAudio31ASR}. These systems provide a foundation for instruction-conditioned audio understanding.

However, autoregressive text generation introduces additional computational overhead and response latency when the required output is a bounded decision. Generated answers may also require parsing and validation before they can be consumed by downstream software. These costs become particularly relevant in applications that process large volumes of audio or require frequent decisions. This motivates an audio decision model that retains natural-language task specification while directly producing outputs in an explicitly defined answer space.

We introduce Qwen-Audio-Jev, a family comprising Qwen-Audio-3.1-Jev and Qwen-Audio-3.1-Jev-Tiny. The former targets high-accuracy complete-clip decisions, while Qwen-Audio-3.1-Jev-Tiny supports both complete-clip and streaming decisions with a compact backbone. Our starting point is an LALM pretrained on a large-scale audio understanding corpus spanning audio captioning, open-ended question answering, audio grounding, and multiple-choice question answering. Each variant augments its pretrained LALM with a dedicated decision head that reads the decoder state at a designated decision-token position. The input places the question first, followed by the candidate descriptions or rating criteria, then the audio and decision token. The head supports three typed readouts: \textsc{Choice} returns a distribution over supplied alternatives, \textsc{Noul} estimates the probability that a binary proposition is true, and \textsc{Score} predicts a distribution over ordered levels and computes its expected rating. Both variants share these decision primitives while using different audio encoders and language-model backbones. This adaptation preserves question-conditioned audio processing while allowing bounded decisions to be read out directly, without autoregressive answer generation.

The interface is inspired by TypeSafe AI's Jev and its typed decision primitives~\citep{almeidaJev}. Related ideas also appear in Google's multimodal Gemini Decisions API~\citep{choi2026geminidecisions}. In the audio domain, AudioJev maps waveforms, questions, and supplied alternatives to candidate distributions and uses an order-consistency objective to reduce sensitivity to candidate permutations~\citep{lv2026audiojev}. Building on this direction, Qwen-Audio-Jev adapted a pre-trained LALM to support all three decision primitives through a dedicated decision head, task-specific training objectives, and joint training on heterogeneous audio datasets.

Training a unified decision model requires accommodating datasets with different annotation schemes, as each example may provide supervision for only a subset of tasks. For example, emotion resources such as MELD~\citep{poria2019meld} and speaker-recognition resources such as VoxCeleb~\citep{nagrani2017voxceleb} provide supervision for different questions about audio. We therefore formulate a common supervision representation that allows each example to contribute only to tasks supported by its available labels. This formulation enables joint training across task-specific datasets without requiring every recording to be annotated for every task, while preserving the distinction between categorical labels, binary relations, and ordered ratings.

Our evaluation covers 32 dataset--task configurations spanning audio question answering, sound and attribute classification, binary judgments, speaker verification, and quality assessment. For \textsc{Choice}, Qwen-Audio-3.1-Jev achieves 79.23\% accuracy on MMAU and 82.38\% on MMSU; for \textsc{Noul}, it achieves 94.20\% accuracy on the VoxCeleb1-O speaker verification benchmark; and for \textsc{Score}, it achieves a normalized mean absolute error of 0.1110 on NISQA overall quality assessment, with Pearson and Spearman correlations of 0.8164 and 0.8113, respectively.

The main contributions of this work are:
\begin{itemize}[label=\textbullet, leftmargin=1em, labelsep=0.4em]
    \item We introduce Qwen-Audio-Jev, a family of two models with a unified interface for selection, binary judgment, and scoring.
    \item We provide a natural-language interface for specifying audio questions and answer spaces, returning structured probabilities or scores directly without autoregressive answer generation or text parsing.
    \item We explore an implementation of streaming decisions in Qwen-Audio-3.1-Jev-Tiny for voice activity detection (VAD) and rolling signal-to-noise ratio (SNR) estimation.
\end{itemize}

\section{System Architecture}
\label{sec:architecture}
\begin{figure}[htbp]
\centering
\includegraphics[width=\linewidth]{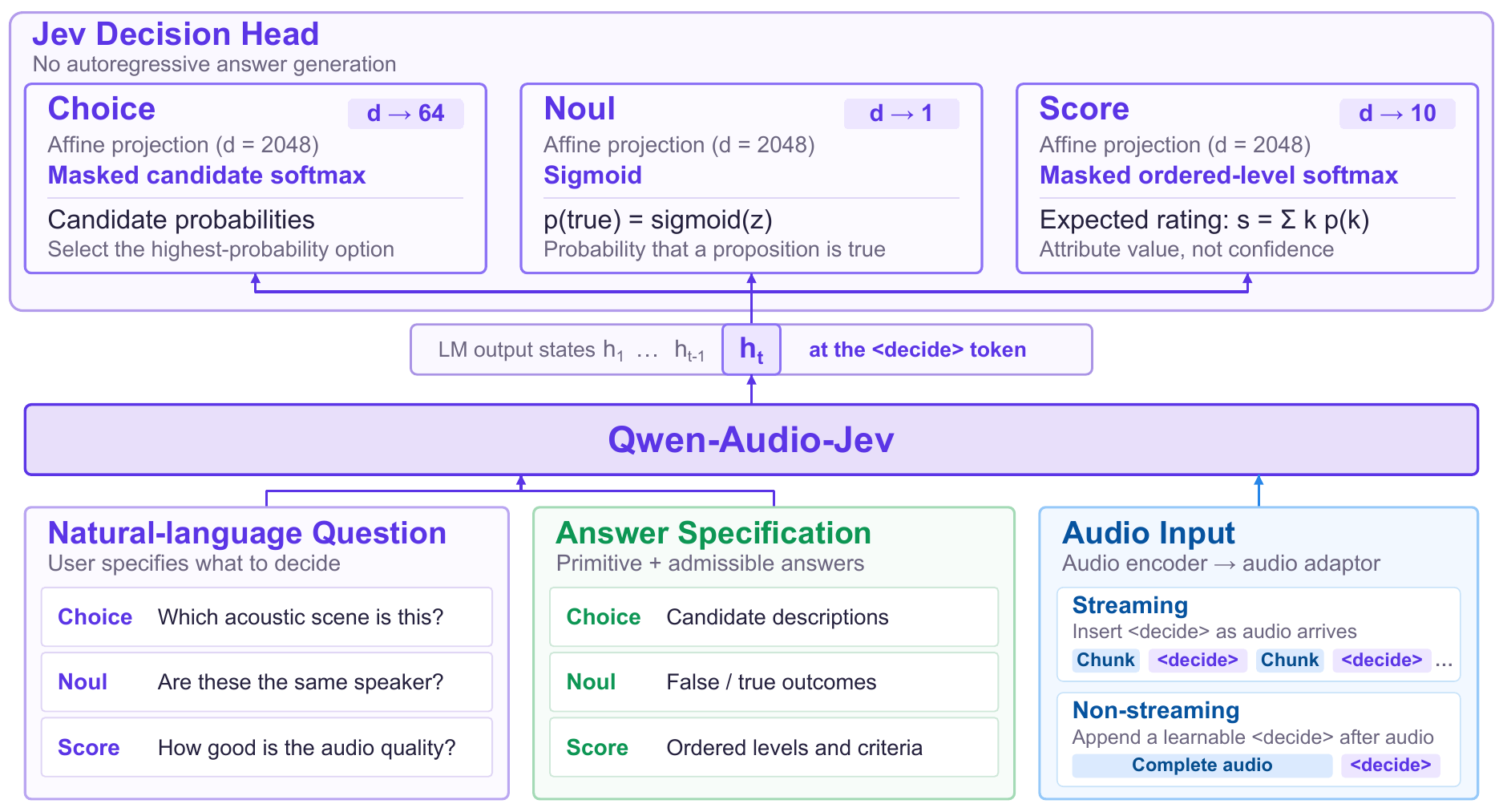}
\caption{Shared architecture of Qwen-Audio-Jev. Inputs follow the order question, candidate descriptions or rating criteria, audio, and decision token. Decoder states at learned decision tokens feed the Choice, Noul, and Score readouts. Complete-clip inputs end with \texttt{<decide>}; Qwen-Audio-3.1-Jev-Tiny's streaming layout places \texttt{<decide\_streaming>} after each audio chunk.}
\label{fig:model-architecture}
\end{figure}

Figure~\ref{fig:model-architecture} presents the shared architecture of Qwen-Audio-Jev. Both variants combine an audio encoder, an audio-to-language adaptor, a language-model decoder, and three typed decision readouts. We first describe the two backbone configurations, then explain the decision-token interface, the complete-clip layout supported by both variants, and the streaming layout supported by Qwen-Audio-3.1-Jev-Tiny.

\subsection{Pretrained LALM Architecture}
\label{sec:backbone-review}
The LALM comprises an audio encoder, an audio-to-language adaptor, and a language-model decoder. For an audio input \(x\), which may contain one or more recordings depending on the task, the representation supplied to the decoder is
\begin{equation}
\mathbf{a}=A_\theta(E_\theta(x)),
\label{eq:audio-representation}
\end{equation}
where \(E_\theta\) and \(A_\theta\) denote the audio encoder and adaptor, respectively. The encoder extracts acoustic features, and the adaptor groups adjacent encoder frames to downsample the sequence before projecting the features to the decoder's hidden dimension. The decoder processes these projected features together with the task instruction, question, and answer specification.

\paragraph{Model variants.}
Qwen-Audio-3.1-Jev pairs an approximately 636M-parameter audio encoder using full attention with the Qwen3.5-35B-A3B mixture-of-experts (MoE) language model. Its decoder has 40 layers, a hidden dimension of 2048, and 256 routed experts, with eight selected per token; the language model has approximately 35B total parameters and 3B active parameters per token. It supports non-streaming inference and targets high-accuracy audio decisions. Qwen-Audio-3.1-Jev-Tiny pairs an approximately 160M-parameter audio encoder using 160\,ms local attention with the dense Qwen3.5-0.8B language model, which has 24 layers and a hidden dimension of 1024. It supports both non-streaming and streaming inference, targeting low-resource, low-latency streaming deployments. Each variant uses an adaptor matched to its decoder width. Table~\ref{tab:model-variants} summarizes the two configurations and their intended use.

\begin{table}[!htbp]
\centering
\small
\setlength{\tabcolsep}{3pt}
\renewcommand{\arraystretch}{1.15}
\caption{Model variants, audio-encoder attention, supported input modes, and intended use. Table and figure labels use Jev for Qwen-Audio-3.1-Jev and Jev-Tiny for Qwen-Audio-3.1-Jev-Tiny.}
\label{tab:model-variants}
\begin{tabularx}{\linewidth}{@{}l >{\raggedright\arraybackslash}p{0.16\linewidth} >{\raggedright\arraybackslash}p{0.19\linewidth} c c >{\raggedright\arraybackslash}X@{}}
\toprule
Variant & \shortstack[l]{Encoder\\Attention} & LLM & \shortstack{Non-\\streaming} & Streaming & Intended Use \\
\midrule
Jev & Full attention & Qwen3.5-\newline 35B-A3B & \(\checkmark\) & \(\times\) & High-accuracy audio decisions \\
Jev-Tiny & Local attention (160\,ms) & Qwen3.5-0.8B & \(\checkmark\) & \(\checkmark\) & Low-resource, low-latency streaming \\
\bottomrule
\end{tabularx}
\end{table}

Before adding the Jev decision head, we perform supervised fine-tuning (SFT) on a large-scale audio understanding corpus that includes audio captioning, open-ended question answering, audio grounding, and multiple-choice question answering. Given audio \(x\), textual instruction and context \(u\), and a reference response \(y\) of length \(L\), the model assigns the response the conditional probability
\begin{equation}
p_\theta(y\mid x,u)=\prod_{\ell=1}^{L}p_\theta(y_\ell\mid y_{<\ell},\mathbf{a},u).
\label{eq:backbone-generation}
\end{equation}
We train with teacher forcing, conditioning the prediction at each response position \(\ell\) on the reference prefix \(y_{<\ell}\) and minimizing the negative log-likelihood of the reference response. Qwen-Audio-Jev uses the resulting decoder representations as inputs to the Jev decision head.

\subsection{Jev Decision-Head Implementation}
\label{sec:decision-head}

Let $q$ be a natural-language question and $c$ its answer
specification, which defines the requested primitive and the
valid answer space. The specification includes candidate
descriptions for \textsc{Choice}, the binary interpretation for
\textsc{Noul}, or ordered rating criteria for \textsc{Score}.
The resulting interface is
\begin{equation}
    f_\theta(x,q,c)=r_\tau,
    \qquad
    \tau\in\{
      \textsc{Choice},\textsc{Noul},\textsc{Score}
    \},
    \label{eq:typed-interface}
\end{equation}
where $\tau$ is the primitive specified by $c$, and $r_\tau$
is its corresponding result.

The input order is question, answer specification, audio,
and decision token. The answer specification lists the
\textsc{Choice} candidates in output-index order, states the
\textsc{Noul} proposition's binary interpretation, or supplies
the ordered \textsc{Score} criteria. The task instruction
accompanies this textual prefix before the audio representations.
In the complete-clip layout, a learnable \texttt{<decide>}
token immediately follows the supplied audio. Its embedding acts as a
decision query, and its final decoder state is used for readout:
\begin{equation}
\begin{aligned}
    \mathbf{H}
      &=D_\theta\!\left(
         \operatorname{Fuse}
         \bigl(\operatorname{Embed}(\operatorname{Serialize}(q,c)),
               \mathbf{a},e_{\mathrm{decide}}
         \bigr)
       \right),\\
    h_\star&=\mathbf{H}_{t_\star}.
\end{aligned}
    \label{eq:decision-state}
\end{equation}
Here, $D_\theta$ is the language-model decoder and
$e_{\mathrm{decide}}$ is the learned decision-token embedding.
$\operatorname{Serialize}(q,c)$ places the question before its
candidate descriptions or rating criteria.
$\operatorname{Fuse}$ then assembles this prefix, adapted
acoustic embeddings, and decision token in that order.
The position $t_\star$ identifies the decision token. Under
causal attention, its state summarizes the preceding task
text and supplied audio evidence. This position is the direct
readout location; no answer tokens are generated to obtain it.

The decision head contains three primitive-specific affine
projections:
\begin{equation}
\begin{aligned}
    z^{\mathrm{Choice}} &= W_C h_\star+b_C,\\
    z^{\mathrm{Noul}}   &= W_N h_\star+b_N,\\
    z^{\mathrm{Score}}  &= W_S h_\star+b_S.
\end{aligned}
    \label{eq:decision-projections}
\end{equation}
In the current implementation,
$W_C\in\mathbb{R}^{64\times d}$,
$W_N\in\mathbb{R}^{1\times d}$, and
$W_S\in\mathbb{R}^{10\times d}$, where $d$ is the decoder hidden dimension: 1024 for Qwen-Audio-3.1-Jev-Tiny and 2048 for Qwen-Audio-3.1-Jev.
The three projections share the same input-conditioned state.
The requested primitive determines which readout supplies the
result and which task objective provides supervision.
Inference requires no autoregressive answer generation.
Although a request may contain multiple questions, the current
implementation processes them separately, requiring a separate
question-conditioned forward pass for each question.

\textsc{Choice} and \textsc{Score} use fixed-capacity index
projections. For a question with fewer candidates or rating
levels than the corresponding capacity, only the valid logits
enter the probability normalization; unused positions are
excluded. The answer specification determines the meanings
of the valid output indices for each question.
Figure~\ref{fig:model-architecture} shows this common readout
structure for both variants; the projection input width changes
with the language-model backbone.

\paragraph{Non-streaming and streaming decision-state readout.}
Both variants retain the question and answer specification
before the audio. They support complete-clip readout: one
\texttt{<decide>} token follows the audio, and its state
produces a clip-level decision. Qwen-Audio-3.1-Jev uses this non-streaming
layout. Qwen-Audio-3.1-Jev-Tiny additionally supports a streaming layout that
interleaves audio chunks
with a separate learned \texttt{<decide\_streaming>} token.
The decoder states at these positions provide successive
decisions from the audio evidence available at each readout.
The two modes use the same primitive-specific heads, while the
distinct decision tokens identify the readout mode. Multiple
streaming readouts answer the same question as evidence arrives;
different questions remain separately conditioned requests.

\subsection{Typed Outputs: Choice, Noul, and Score}
\label{sec:typed-outputs}

The three readouts follow Jev's typed-question
terminology. Their answer
spaces and output transformations are defined below.

\paragraph{Choice: A Distribution over Supplied Alternatives.}
A \textsc{Choice} specification supplies an ordered collection
of $N$ candidate descriptions, with $2\le N\le64$ in the current
implementation. The candidate descriptions follow the question
and precede the audio, in the same order as the output indices.
The readout uses the first $N$ \textsc{Choice} logits:
\begin{equation}
    p_i
      =\frac{\exp(z_i^{\mathrm{Choice}})}
             {\sum_{j=1}^{N}\exp(z_j^{\mathrm{Choice}})},
    \qquad
    \hat{\imath}
      =\arg\max_{i\in\{1,\ldots,N\}}p_i.
    \label{eq:choice-output}
\end{equation}
The result contains the candidate probabilities and the option
identified by $\hat{\imath}$.

This implementation scores candidate positions from the shared
decision state rather than assigning a fixed global class
meaning to each output position. For example, the same output
indices may refer to acoustic scenes in one question and emotion
categories in another, according to the supplied descriptions.

\paragraph{Noul: The Probability That a Proposition Is True.}
A \textsc{Noul} question specifies a binary proposition about
the audio, such as whether two recordings contain the same
speaker. Its scalar logit is transformed into
\begin{equation}
    p_{\mathrm{true}}
      =\sigma(z^{\mathrm{Noul}})
      =\frac{1}{1+\exp(-z^{\mathrm{Noul}})}.
    \label{eq:noul-output}
\end{equation}
The native output is a probability in $[0,1]$.
Applications may obtain a binary decision by comparing it
with a task-specific threshold. The threshold is a downstream
decision rule rather than an additional model output.

\paragraph{Score: An Expectation over Ordered Rating Levels.}
A \textsc{Score} specification defines $K$ ordered levels,
with $2\le K\le10$ in the current implementation.
The first $K$ \textsc{Score} logits produce
\begin{equation}
    p_k
      =\frac{\exp(z_k^{\mathrm{Score}})}
             {\sum_{j=0}^{K-1}\exp(z_j^{\mathrm{Score}})},
    \qquad k=0,\ldots,K-1.
    \label{eq:score-distribution}
\end{equation}
The native score is the probability-weighted level index:
\begin{equation}
    \hat{s}_{\mathrm{idx}}
      =\sum_{k=0}^{K-1}k p_k,
    \qquad
    \hat{s}_{\mathrm{norm}}
      =\frac{\hat{s}_{\mathrm{idx}}}{K-1}.
    \label{eq:score}
\end{equation}
The model thus predicts a distribution over the declared
levels, from which a continuous expected level index is
computed. The native and normalized values lie in $[0,K-1]$
and $[0,1]$, respectively.

For equally spaced application levels spanning $[a,b]$,
the expected rating in application units is
\begin{equation}
    \hat{s}_{[a,b]}
      =a+(b-a)\hat{s}_{\mathrm{norm}}.
    \label{eq:score-scale-mapping}
\end{equation}
For non-equally spaced numerical levels $v_0,\ldots,v_{K-1}$,
the corresponding expected rating is
$\sum_{k=0}^{K-1}v_k p_k$.
Such scale conversion is separate from the native index-based
\textsc{Score} output. Predictions and reference ratings must
use the same mapping in evaluation.

\begin{table}[htbp]
\centering
\small
\setlength{\tabcolsep}{4pt}
\renewcommand{\arraystretch}{1.18}
\caption{Answer spaces and native outputs of the three decision
readouts in the current implementation.}
\label{tab:primitives}

\begin{tabularx}{\linewidth}{
    @{}
    >{\raggedright\arraybackslash}p{0.15\linewidth}
    >{\hsize=0.8\hsize\linewidth=\hsize
      \raggedright\arraybackslash}X
    >{\hsize=1.2\hsize\linewidth=\hsize
      \raggedright\arraybackslash}X
    @{}
}
\toprule
\textbf{Type} & \textbf{Answer space} & \textbf{Native output} \\
\midrule
\textsc{Choice}
& 2--64 ordered candidate descriptions
& Candidate probabilities and the selected option \\[2pt]

\textsc{Noul}
& A binary proposition
& Probability that the proposition is true \\[2pt]

\textsc{Score}
& 2--10 ordered rating levels
& Level probabilities and the expected level index \\
\bottomrule
\end{tabularx}
\end{table}

Table~\ref{tab:primitives} summarizes the answer spaces and
native outputs of the three decision readouts. 
These outputs provide a common interface for categorical
selection, binary judgment, and ordered scoring.

\section{Training}
\label{sec:training}

\begin{figure}[htbp]
\centering
\includegraphics[width=\linewidth]{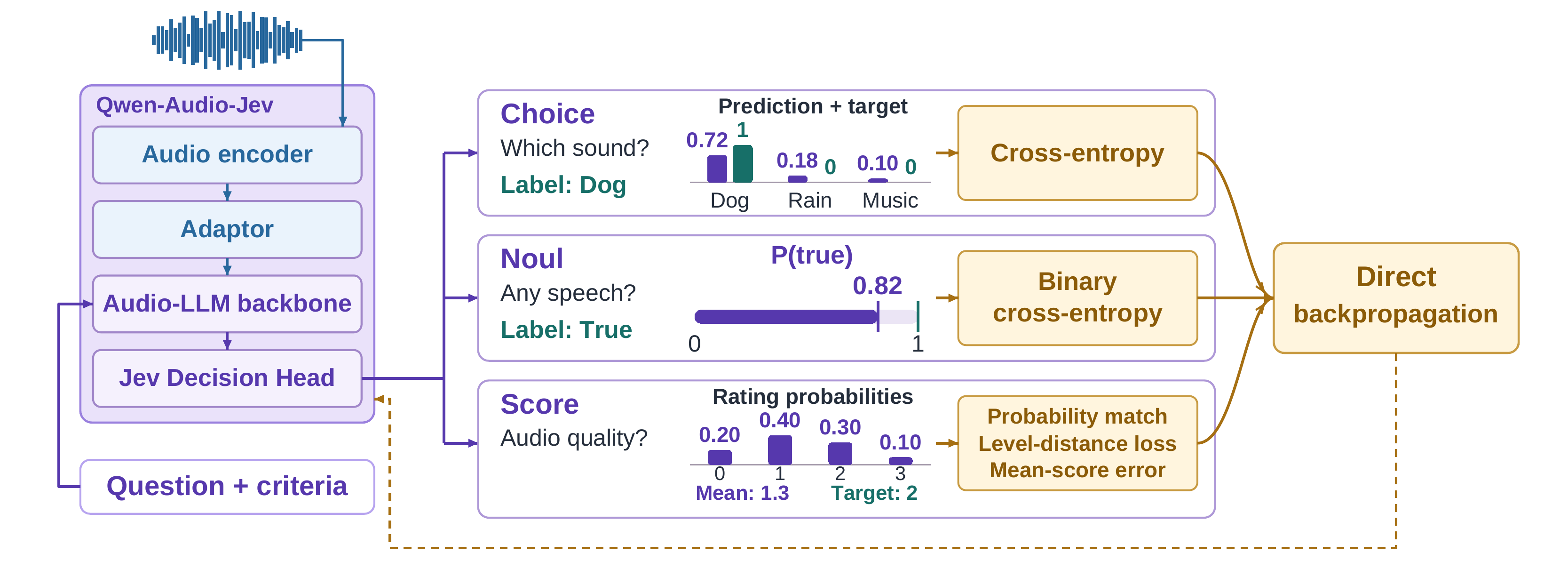}
\caption{Supervised fine-tuning of Qwen-Audio-Jev. SFT compares predicted distributions with observed labels and directly backpropagates task-specific losses. Choice, Noul, and Score are alternative question-conditioned tasks; only the requested primitive is active.}
\label{fig:training-process}
\end{figure}

We train the three decision primitives through supervised fine-tuning (SFT) on heterogeneous task annotations, as illustrated in Figure~\ref{fig:training-process}.

The supervised stage jointly optimizes the audio encoder,
audio-to-language adaptor, language-model backbone, and Jev
decision head. Each question-level training example activates
the objective associated with its requested decision primitive,
while the remaining readouts receive no task-specific supervision.
Qwen-Audio-3.1-Jev-Tiny is jointly trained on non-streaming and streaming examples during SFT.

\subsection{Training Formulation and Supervision}

Multi-task audio pretraining and audio--question--answer
instruction data have been widely used to improve general
audio understanding~\citep{ghosh2024gama,gong2024ltu,xu2025qwen3omni}.
We organize all training examples into a unified
audio--question--criteria--target format. Each example $i$
contains the audio input, a question, an active primitive,
the corresponding answer specification, and an observed target.

Categorical labels are used to construct \textsc{Choice}
targets, binary judgments to construct \textsc{Noul} targets,
and audio-grounded annotations with an explicit ordinal
structure to construct \textsc{Score} targets.
An individual audio recording need not be annotated for all
three primitives. Available annotations are used to construct
question-level examples, each of which supplies an observed
target for its active primitive. Missing annotations provide
no supervision.
All primitive-specific losses are computed from the
corresponding decision-head logits.

\subsection{Primitive-Specific Objectives}
\label{sec:task-losses}

\paragraph{Choice: Categorical Cross-Entropy.}

Let $\mathcal{C}_i$ be the set of valid candidate indices,
$z_{ik}$ the logit for candidate $k$, and $t_{ik}$ its target
probability, with $t_{ik}\geq0$ and
$\sum_{k\in\mathcal{C}_i}t_{ik}=1$. We define
\begin{align}
    p_{ik}
    &= \frac{\exp(z_{ik})}
    {\sum_{j\in\mathcal{C}_i}\exp(z_{ij})},
    & k &\in\mathcal{C}_i,
    \label{eq:choice-prob}\\
    \ell_i^{\mathrm{choice}}
    &= -\sum_{k\in\mathcal{C}_i}t_{ik}\log p_{ik}.
    \label{eq:choice-loss}
\end{align}
Unused output positions are excluded from the softmax and carry
zero target mass. A hard class label $y_i$ gives the one-hot
target $t_{ik}=\mathbf{1}[k=y_i]$, reducing the loss to
$-\log p_{i,y_i}$. Soft targets are used when annotation
distributions are available. This objective covers nominal
categories such as emotion, scene, instrument, and sound class,
as well as multiple-choice audio questions.

During training, \textsc{Choice} alternatives are randomly
permuted, with their descriptions and target probabilities
reordered accordingly. This augmentation discourages the
model from relying on fixed answer positions.

\paragraph{Noul: Binary Cross-Entropy.}

For a binary proposition, $p_i=\sigma(z_i)$ denotes the
probability that the proposition is true, and
$y_i\in\{0,1\}$ is the observed outcome. The per-question
objective is
\begin{equation}
    \ell_i^{\mathrm{noul}}
    = -y_i\log p_i-(1-y_i)\log(1-p_i).
    \label{eq:noul-loss}
\end{equation}
The same objective also supports soft targets $y_i\in[0,1]$.
Speaker verification and yes/no audio judgments are trained
using this objective.

\paragraph{Score: Distribution, Ordering, and Expected Value.}

For $K_i$ ordered levels indexed from $0$ to $K_i-1$,
the predicted probabilities are normalized over the first
$K_i$ valid \textsc{Score} logits:
\begin{equation}
    p_{ik}
    = \frac{\exp(z_{ik})}
    {\sum_{j=0}^{K_i-1}\exp(z_{ij})},
    \qquad k=0,\ldots,K_i-1.
    \label{eq:score-train-prob}
\end{equation}
Let $t_{ik}$ denote the target level probability, with
$t_{ik}\geq0$ and $\sum_{k=0}^{K_i-1}t_{ik}=1$.

The expected level indices and cumulative probabilities are
\begin{align}
    \hat{s}_i &= \sum_{k=0}^{K_i-1} k p_{ik},
    & s_i^\star &= \sum_{k=0}^{K_i-1} k t_{ik},\\
    P_{im} &= \sum_{k=0}^{m}p_{ik},
    & T_{im} &= \sum_{k=0}^{m}t_{ik},
\end{align}
where $m=0,\ldots,K_i-2$.
The three loss components are
\begin{align}
    C_i
    &= -\sum_{k=0}^{K_i-1}t_{ik}\log p_{ik},
    \label{eq:score-ce}\\
    D_i
    &= \frac{1}{K_i-1}
       \sum_{m=0}^{K_i-2}(P_{im}-T_{im})^2,
    \label{eq:score-cdf}\\
    H(e)
    &= \begin{cases}
        \tfrac{1}{2}e^2, & |e|<1,\\
        |e|-\tfrac{1}{2}, & |e|\geq 1.
    \end{cases}
    \label{eq:smooth-l1}
\end{align}
The combined per-question loss is
\begin{equation}
    \ell_i^{\mathrm{score}}
    = C_i+\lambda_{\mathrm{cdf}}D_i
      +\lambda_{\mathrm{mean}}H(\hat{s}_i-s_i^\star).
    \label{eq:score-loss}
\end{equation}
Cross-entropy matches the target level distribution,
the CDF term penalizes ordinal displacement, and the SmoothL1
term penalizes errors in the expected level index on the
native $[0,K_i-1]$ scale.
We set $\lambda_{\mathrm{cdf}}=0.25$ and
$\lambda_{\mathrm{mean}}=0.10$ in all SFT experiments.

\subsection{Overall Objective and Task Sampling}
\label{sec:loss-reduction}

Each question-level example contributes the loss associated
with its active primitive. Let $\tau_i$ denote the active
primitive of example $i$. Its decision loss is
\begin{equation}
    \ell_i^{\mathrm{dec}}=\ell_i^{\tau_i}.
    \label{eq:decision-reduction}
\end{equation}
We denote the decision loss aggregated over sampled
question-level examples by $\mathcal{L}_{\mathrm{dec}}$.

The relative frequency of the three decision primitives is
controlled by their sampling distribution in the training
stream.

The SFT objective depends on the backbone variant:
\begin{equation}
    \mathcal{L}_{\mathrm{SFT}}
      =\begin{cases}
        \mathcal{L}_{\mathrm{dec}}
        +\lambda_{\mathrm{lb}}\mathcal{L}_{\mathrm{lb}}^{\mathrm{Jev}}
        +\mathcal{L}_{\mathrm{rz}}, & \text{Jev},\\
        \mathcal{L}_{\mathrm{dec}}, & \text{Jev-Tiny}.
      \end{cases}
    \label{eq:sft-total}
\end{equation}
Here, $\mathcal{L}_{\mathrm{lb}}^{\mathrm{Jev}}$ denotes the
aggregated load-balancing loss of the language-model backbone,
$\lambda_{\mathrm{lb}}$ is its auxiliary coefficient, and
$\mathcal{L}_{\mathrm{rz}}$ denotes the router z-loss term
returned by the backbone.
These MoE auxiliary terms apply only to Qwen-Audio-3.1-Jev. Qwen-Audio-3.1-Jev-Tiny uses a dense
backbone and does not require either the load-balancing loss or
the router z-loss.

\section{Evaluation}

\label{sec:evaluation}

\subsection{Datasets}

We evaluate Qwen-Audio-3.1-Jev and Qwen-Audio-3.1-Jev-Tiny on complete-clip decisions, and the latter on streaming decisions. The complete-clip suite contains 32 dataset--task configurations: 15 for \textsc{Choice}, seven for \textsc{Noul}, and ten for \textsc{Score}, covering acoustic classification, audio question answering, binary judgments, and human-rated acoustic attributes. The streaming suite covers voice activity detection and rolling SNR estimation.

Table~\ref{tab:jev-eval-datasets} groups the evaluation suite by decision type. Note that Smart Turn evaluates turn completion on 31,527 clips using up to the final eight seconds of each clip. MMAU-mini is the 1,000-question mini subset with public answers; MMAU is the full 10,000-question benchmark, including a further 9,000 questions with hidden answers. We submit predictions for the latter to the official evaluation service\footnote{Official MMAU evaluation service: \url{https://huggingface.co/spaces/sonalkum/MMAU-Eval}.} and compute overall MMAU accuracy as the combined number of correct predictions across both subsets divided by 10,000.

\begin{table}[!htbp]
\centering
\small
\caption{Core Evaluation Datasets}
\label{tab:jev-eval-datasets}
\begin{tabularx}{\linewidth}{@{}l X@{}}
\toprule
Primitive & Datasets \\
\midrule
\textsc{Choice} & CochlScene~\citep{jeong2022cochlscene}, NSynth~\citep{engel2017nsynth}, NonSpeech7k~\citep{rashid2023nonspeech7k}, UrbanSound8K, MELD~\citep{poria2019meld}, DiDiSpeech~\citep{guo2020didispeech}, LibriSpeech test-clean, MINDS-14, Speech Commands v2, FLEURS, Samrómur, MMAU-mini and MMAU~\citep{sakshi2024mmau}, MMSU~\citep{wang2026mmsu}, MMAR~\citep{ma2025mmar}. \\
\addlinespace
\textsc{Noul} & VoxCeleb1-O~\citep{nagrani2017voxceleb}, Clotho-AQA~\citep{lipping2022clothoaqa}, PodSarc~\citep{li2025podsarc}, In-the-Wild~\citep{muller2022inthewild}, AMI test, Smart Turn. \\
\addlinespace
\textsc{Score} & NISQA LiveTalk~\citep{mittag2021nisqa}, VoiceMOS/BVCC~\citep{cooper2021bvcc}, SpeechOcean762, TMHINT-QI. \\
\addlinespace
Streaming decisions & AISHELL-4~\citep{fu2021aishell4}, Rolling SNR. \\
\bottomrule
\end{tabularx}
\end{table}

\FloatBarrier

\subsection{Choice}

\label{sec:eval-choice}

\paragraph{Metrics.}

We measure top-1 accuracy and compute macro-F1 for tasks with a fixed label set. For the other Choice tasks, selections matching the reference option text are counted as correct. We assess answer-position stability on 1,000 MMAU-mini, 5,000 MMSU, and 1,000 MMAR questions by keeping each question and its real options fixed while varying their order. For a question with $K$ real options, we evaluate $K$ cyclic reorderings of those options. We also pad the option list to 64 with \texttt{None} and rotate the correct answer through every slot. The primary 64-slot score counts filler selections as errors, and a secondary score masks fillers. We compute per-position and mean accuracy, and the mean fraction of rotations that select the same modal real answer for each question.

\paragraph{Results.}

The Qwen-Audio-3.1-Jev model is most reliable when decisions depend on well-defined acoustic or lexical cues, as illustrated by language identification, gender classification, short-command recognition, and nonverbal vocal events (Table~\ref{tab:classification}). Emotion and age judgments are less consistent. On MELD, the gap between accuracy and macro-F1 indicates uneven performance across emotion categories, rather than uniformly strong recognition. Environmental sound classification also varies substantially across datasets, showing that success on one sound taxonomy does not transfer uniformly to others.

For question-conditioned audio understanding, Qwen-Audio-3.1-Jev achieves 79.23\% on MMAU and 82.38\% on MMSU. Its lower MMAR accuracy points to a remaining gap on questions that combine multiple audio cues and relationships. These results distinguish robust recognition of salient attributes from the more demanding integration of context, affect, and acoustic events; the model produces its final decision directly in all cases.

Qwen-Audio-3.1-Jev-Tiny achieves 71.70\% on MMAU-mini and 71.79\% on MMAU, alongside 66.58\% on MMSU and 52.90\% on MMAR. It remains close to Qwen-Audio-3.1-Jev on NonSpeech7k and gender classification, but has larger gaps on language identification and intent recognition. It therefore preserves several salient acoustic attributes while providing weaker question-conditioned and language-dependent decisions than Qwen-Audio-3.1-Jev.

\begin{table}[!htbp]
\centering
\small
\caption{Choice accuracy and macro-F1 across acoustic classification and audio question-answering benchmarks (\%). MMAU denotes the full 10,000-question benchmark.}
\label{tab:classification}
\begin{tabular}{@{}lrrrr@{}}
\toprule
 & \multicolumn{2}{c}{Accuracy $\uparrow$} & \multicolumn{2}{c}{Macro-F1 $\uparrow$} \\
\cmidrule(lr){2-3}\cmidrule(lr){4-5}
Benchmark & Jev & Jev-Tiny & Jev & Jev-Tiny \\
\midrule
CochlScene & 79.16 & 78.24 & 78.96 & 78.08 \\
NSynth & 73.44 & 77.20 & 73.48 & 78.14 \\
NonSpeech7k & 92.69 & 92.41 & 92.35 & 92.55 \\
DiDiSpeech & 100.00 & 100.00 & 100.00 & 100.00 \\
MELD & 56.85 & 53.93 & 38.84 & 33.24 \\
MMAU-mini & 78.90 & 71.70 & -- & -- \\
MMAU & 79.23 & 71.79 & -- & -- \\
MMSU & 82.38 & 66.58 & -- & -- \\
MMAR & 67.70 & 52.90 & -- & -- \\
UrbanSound8K & 64.55 & 65.17 & 59.80 & 59.51 \\
LibriSpeech gender & 98.93 & 98.59 & 98.93 & 98.58 \\
MINDS-14 & 82.88 & 46.41 & 83.33 & 44.00 \\
Speech Commands v2 & 93.48 & 83.15 & 93.30 & 78.88 \\
FLEURS (20 languages) & 99.10 & 40.10 & 99.10 & 36.34 \\
Samrómur age & 67.50 & 55.50 & 68.17 & 52.91 \\
\bottomrule
\end{tabular}
\end{table}

\begin{figure}[!htbp]
\centering
\includegraphics[width=0.82\linewidth,trim=0 18 0 8,clip]{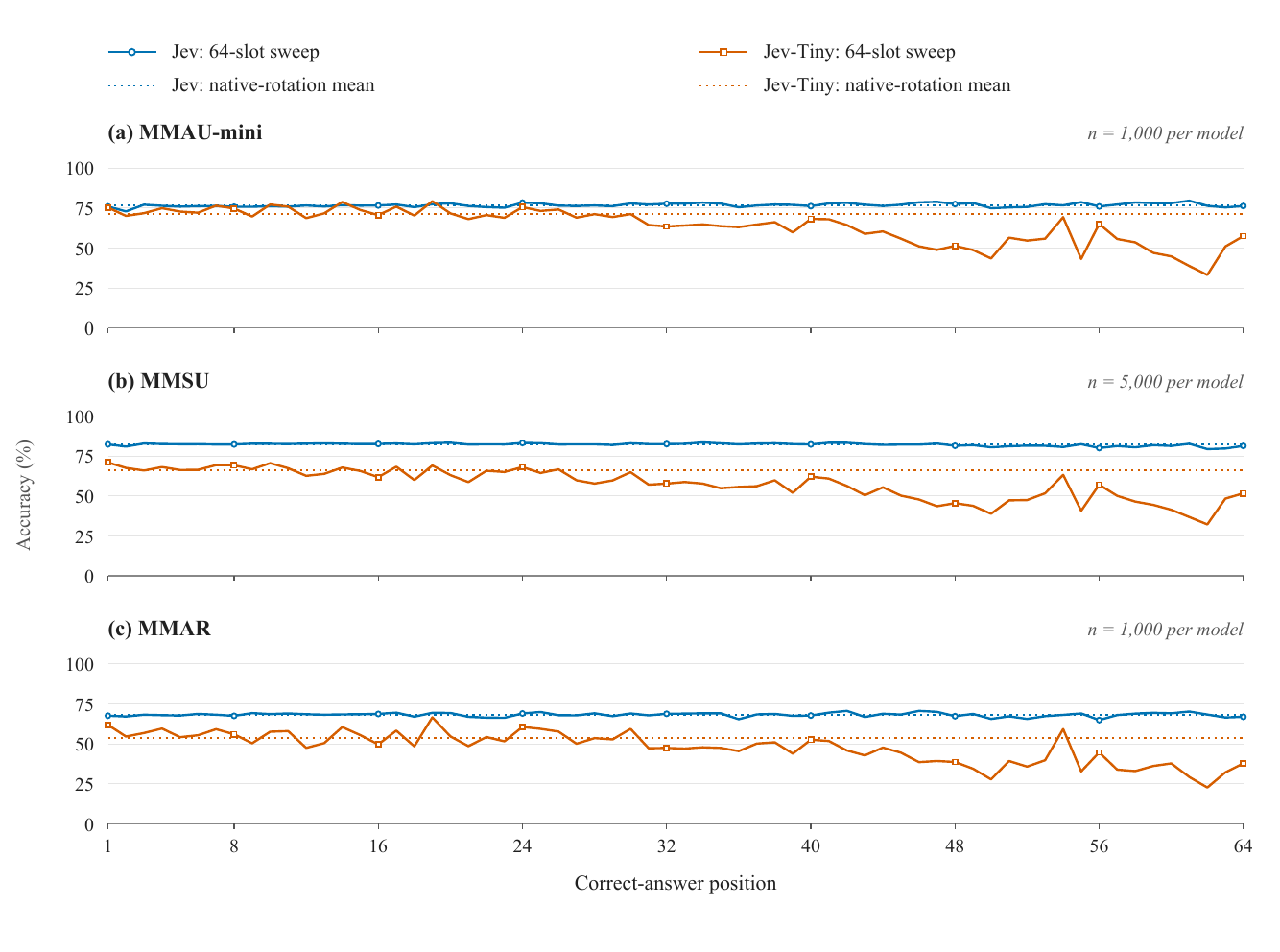}
\caption{Answer-position stability of the Qwen-Audio-Jev models. Solid lines show raw accuracy in the 64-slot sweep; dotted lines show native-rotation means. Selecting \texttt{None} counts as an error.}
\label{fig:choice-position-accuracy}
\end{figure}

\paragraph{Answer-Position Stability.}
Figure~\ref{fig:choice-position-accuracy} compares the two Qwen-Audio-Jev models across all 64 correct-answer positions. Qwen-Audio-3.1-Jev remains comparatively stable: its mean accuracy differs by less than 0.3 percentage points between native and 64-slot rotations on all three benchmarks, while semantic consistency remains above 91\% in both conditions (Table~\ref{tab:choice-position}). Qwen-Audio-3.1-Jev-Tiny's mean accuracy decreases by 7.00, 8.52, and 5.88 percentage points on MMAU-mini, MMSU, and MMAR, respectively. Its semantic consistency also decreases in the 64-slot sweep, indicating greater sensitivity to answer position under this protocol.

\FloatBarrier

\begin{table}[!htbp]
\centering
\small
\setlength{\tabcolsep}{4pt}
\caption{Answer-position stability of the Qwen-Audio-Jev models (\%) under native-option rotations and the 64-slot sweep. The 64-slot columns count filler selections as errors.}
\label{tab:choice-position}
\begin{tabular}{@{}llrrrr@{}}
\toprule
 & & \multicolumn{2}{c}{Mean accuracy} & \multicolumn{2}{c}{Semantic consistency} \\
\cmidrule(lr){3-4}\cmidrule(lr){5-6}
Benchmark & Model & Native & 64 slots & Native & 64 slots \\
\midrule
MMAU-mini & Jev & 76.49 & 76.74 & 92.43 & 93.24 \\
 & Jev-Tiny & 71.01 & 64.01 & 91.32 & 78.06 \\
\addlinespace[2pt]
MMSU & Jev & 82.23 & 82.07 & 95.11 & 95.26 \\
 & Jev-Tiny & 65.99 & 57.47 & 88.47 & 74.63 \\
\addlinespace[2pt]
MMAR & Jev & 68.04 & 68.08 & 91.68 & 91.75 \\
 & Jev-Tiny & 53.75 & 47.87 & 88.81 & 73.45 \\
\bottomrule
\end{tabular}
\end{table}

\subsection{Noul}

\label{sec:eval-noul}

\paragraph{Metrics.}

For complete-clip tasks, we compute accuracy, balanced accuracy, F1, and false-positive and false-negative rates from decisions at a probability threshold of 0.5. ROC AUC and equal error rate use continuous predicted probabilities and summarize discrimination across thresholds. The binary Brier score is the mean squared error of the predicted probability against the binary label. VoxCeleb1-O also uses minimum detection cost at target priors of 0.01 and 0.05. Streaming thresholds and metrics follow Section~\ref{sec:streaming-benchmarks}.

\paragraph{Results.}

Binary judgments vary substantially across tasks (Table~\ref{tab:noul-results}). Speaker verification, speech activity detection, and turn completion combine high accuracy with strong discrimination across thresholds. Clotho-AQA requires the proposition to be grounded in the recording rather than inferred from a single salient cue.

The In-the-Wild results expose a different limitation. At the fixed 0.5 threshold, the false-negative rate is 59.31\%, while ROC AUC is 72.55\%; both the operating-point decisions and cross-threshold ranking leave substantial room for improvement. Smart Turn presents a more balanced operating point, with false-positive and false-negative rates of 3.98\% and 3.24\%. These contrasts show why thresholded accuracy, discrimination, and probability error provide complementary views of the same Noul interface.

Qwen-Audio-3.1-Jev-Tiny reaches 93.23\% accuracy on VoxCeleb1-O and 95.49\% on Smart Turn. Its lower Clotho-AQA accuracy of 69.67\% highlights remaining difficulty with audio-grounded propositions.

\begin{table}[!htbp]
\centering
\small
\setlength{\tabcolsep}{2pt}
\caption{Noul performance across speaker verification, audio-grounded judgments, and turn completion. Accuracy, ROC AUC, and EER are percentages; Brier uses the 0--1 scale.}
\label{tab:noul-results}
\begin{tabular}{@{}lrrrrrrrr@{}}
\toprule
 & \multicolumn{2}{c}{Accuracy $\uparrow$} & \multicolumn{2}{c}{ROC AUC $\uparrow$} & \multicolumn{2}{c}{EER $\downarrow$} & \multicolumn{2}{c}{Brier $\downarrow$} \\
\cmidrule(lr){2-3}\cmidrule(lr){4-5}\cmidrule(lr){6-7}\cmidrule(lr){8-9}
Benchmark & Jev & Jev-Tiny & Jev & Jev-Tiny & Jev & Jev-Tiny & Jev & Jev-Tiny \\
\midrule
VoxCeleb1-O & 94.20 & 93.23 & 98.41 & 98.01 & 5.66 & 6.85 & 0.0471 & 0.0532 \\
Clotho-AQA & 77.14 & 69.67 & 84.14 & 77.00 & 22.36 & 29.28 & 0.1759 & 0.2344 \\
PodSarc & 99.86 & 99.86 & 100.00 & 99.98 & 0.13 & 0.23 & 0.0010 & 0.0014 \\
In-the-Wild & 69.04 & 69.46 & 72.55 & 78.70 & 33.36 & 27.94 & 0.2276 & 0.2029 \\
AMI speech activity & 93.70 & 91.30 & 98.49 & 96.36 & 5.20 & 8.40 & 0.0474 & 0.0743 \\
AMI speech overlap & 89.80 & 86.90 & 94.68 & 94.83 & 10.00 & 11.40 & 0.0910 & 0.1163 \\
Smart Turn & 96.39 & 95.49 & 99.20 & 99.00 & 3.65 & 4.57 & 0.0314 & 0.0379 \\
\bottomrule
\end{tabular}
\end{table}

\subsection{Score}

\label{sec:eval-score}

\paragraph{Metrics.}

We compute MAE and RMSE on ratings normalized to $[0,1]$ and in the original rating units. We convert normalized errors to the original units by multiplying by the range of the rating scale. Pearson correlation measures linear association with human ratings, and Spearman correlation measures rank agreement.

\paragraph{Results.}

The Score readout aligns more closely with human judgments of recording quality than with detailed assessments of pronunciation and delivery (Table~\ref{tab:score-results}). VoiceMOS/BVCC shows the strongest overall agreement, while NISQA overall quality reaches a normalized MAE of 0.1110, equivalent to 0.4439 MOS points. The NISQA dimensions are not interchangeable: discontinuity has strong rank agreement despite larger absolute errors than coloration, illustrating the distinction between ordering recordings correctly and assigning accurate rating values.

SpeechOcean762 is more challenging across pronunciation accuracy, fluency, and prosody, with weaker correlations than the recording-quality benchmarks. TMHINT-QI falls between these groups in rank agreement. The shared scoring interface therefore supports several kinds of human-rated attributes, but its current strength lies in perceptual recording quality; fine-grained delivery and pronunciation judgments remain priorities for further training.

Qwen-Audio-3.1-Jev-Tiny's normalized MAE is 0.1090 on NISQA overall quality and 0.0918 on VoiceMOS/BVCC, with Pearson correlations of 0.8057 and 0.8785. Both errors are lower than Qwen-Audio-3.1-Jev's in this evaluation. On SpeechOcean762, lower errors coexist with modest correlations; TMHINT-QI instead shows a larger error despite stronger rank agreement. These results reinforce the need to report error and correlation together.

\begin{table}[!htbp]
\centering
\small
\setlength{\tabcolsep}{2pt}
\caption{Score prediction error and agreement with human ratings across recording quality, pronunciation, and intelligibility. MAE and RMSE use normalized ratings in $[0,1]$.}
\label{tab:score-results}
\begin{tabular}{@{}lrrrrrrrr@{}}
\toprule
 & \multicolumn{2}{c}{MAE $\downarrow$} & \multicolumn{2}{c}{RMSE $\downarrow$} & \multicolumn{2}{c}{Pearson $\uparrow$} & \multicolumn{2}{c}{Spearman $\uparrow$} \\
\cmidrule(lr){2-3}\cmidrule(lr){4-5}\cmidrule(lr){6-7}\cmidrule(lr){8-9}
Benchmark & Jev & Jev-Tiny & Jev & Jev-Tiny & Jev & Jev-Tiny & Jev & Jev-Tiny \\
\midrule
NISQA overall & 0.1110 & 0.1090 & 0.1405 & 0.1402 & 0.8164 & 0.8057 & 0.8113 & 0.7824 \\
NISQA noisiness & 0.1464 & 0.1468 & 0.1898 & 0.1856 & 0.6689 & 0.7253 & 0.7074 & 0.7581 \\
NISQA coloration & 0.0994 & 0.1035 & 0.1286 & 0.1324 & 0.7823 & 0.7684 & 0.7893 & 0.7413 \\
NISQA discontinuity & 0.1202 & 0.1040 & 0.1523 & 0.1333 & 0.8513 & 0.8617 & 0.8002 & 0.8328 \\
NISQA loudness & 0.1472 & 0.1443 & 0.1749 & 0.1731 & 0.7355 & 0.7153 & 0.6944 & 0.6963 \\
VoiceMOS/BVCC & 0.0969 & 0.0918 & 0.1222 & 0.1178 & 0.8673 & 0.8785 & 0.8616 & 0.8734 \\
SpeechOcean762 accuracy & 0.2164 & 0.1572 & 0.2369 & 0.1810 & 0.4117 & 0.3291 & 0.4266 & 0.3499 \\
SpeechOcean762 fluency & 0.1612 & 0.0979 & 0.1866 & 0.1288 & 0.3906 & 0.4299 & 0.3747 & 0.4654 \\
SpeechOcean762 prosody & 0.1885 & 0.1212 & 0.2143 & 0.1445 & 0.3249 & 0.4025 & 0.3110 & 0.4196 \\
TMHINT-QI intelligibility & 0.1959 & 0.2340 & 0.2315 & 0.2813 & 0.6085 & 0.6415 & 0.6288 & 0.6767 \\
\bottomrule
\end{tabular}
\end{table}

\subsection{Streaming Decisions}

\label{sec:streaming-benchmarks}
We evaluate Qwen-Audio-3.1-Jev-Tiny's streaming decisions on voice activity detection and rolling SNR estimation (Table~\ref{tab:streaming-benchmarks}). Audio is sampled at 16\,kHz, and decisions are read at 160\,ms intervals. The question and answer specification precede the audio; each audio chunk is followed by a \texttt{<decide\_streaming>} token. Streaming evaluation replays recordings to simulate audio arriving over time. Each decision uses only the current and preceding audio, and each evaluation window is processed independently. Reference masks identify the frames included in scoring.

\paragraph{Noul: Voice Activity Detection.}
We derive frame-level speech-presence labels from AISHELL-4 meeting annotations~\citep{fu2021aishell4}. The question asks whether human speech is active at the current time, and the \textsc{Noul} output gives the corresponding probability. We report frame-level accuracy, F1, recall, ROC AUC, and EER. The decision threshold is selected by validation F1 and fixed for testing; ROC AUC and EER use the continuous probabilities.

\paragraph{Score: Rolling SNR Estimation.}
The Rolling SNR benchmark assigns each decision a speech signal-to-noise ratio level for the preceding 1.28\,s window. The answer specification defines ten ordered levels, indexed from 0 to 9, with boundaries at $-10,-5,0,5,10,15,20,25,30$\,dB. For this ordinal benchmark, the most probable level is compared with the reference level. We report mean absolute error in level units, exact-level accuracy, and the fraction of decisions within one level of the reference. This discrete benchmark readout complements the expected-value \textsc{Score} interface used for continuous ratings.

\begin{table}[!htbp]
\centering
\small
\caption{Streaming evaluation benchmarks and test-set coverage. Readouts follow a 160\,ms decision grid.}
\label{tab:streaming-benchmarks}
\begin{tabularx}{\linewidth}{@{}l X r r@{}}
\toprule
Primitive & Benchmark / task & Audio segments & Scored readouts \\
\midrule
\textsc{Noul} & AISHELL-4 / voice activity detection & 1,441 & 286,310 \\
\textsc{Score} & Rolling SNR / ten-level SNR estimation & 360 & 29,736 \\
\bottomrule
\end{tabularx}
\end{table}

\paragraph{Results.}
Table~\ref{tab:streaming-results} reports Qwen-Audio-3.1-Jev-Tiny's streaming results. AISHELL-4 reaches 95.67\% F1 and 97.35\% recall. Rolling SNR has a mean absolute error of 0.6348 levels, with 88.79\% of decisions within one level of the reference. These measurements are based on recording replay and do not measure real-time inference latency.

\begin{table}[!htbp]
\centering
\small
\caption{Jev-Tiny streaming results. All values are percentages except ordinal MAE. The VAD threshold is 0.20, selected on validation data.}
\label{tab:streaming-results}
\begin{tabularx}{\linewidth}{@{}l X r@{}}
\toprule
Benchmark & Metric & Jev-Tiny \\
\midrule
AISHELL-4 & Accuracy & 92.03 \\
 & F1 & 95.67 \\
 & Recall & 97.35 \\
 & ROC AUC & 89.88 \\
 & EER & 18.35 \\
\addlinespace
Rolling SNR & Ordinal MAE (levels) & 0.6348 \\
 & Exact-level accuracy & 53.13 \\
 & Within one level & 88.79 \\
\bottomrule
\end{tabularx}
\end{table}

\section{Conclusion}
\label{sec:conclusion}

We present Qwen-Audio-Jev, a family of two models that augment pretrained LALMs with a shared interface for \textsc{Choice}, \textsc{Noul}, and \textsc{Score}. Both produce question-conditioned decisions directly from a decision-token state without autoregressive answer generation; Qwen-Audio-3.1-Jev-Tiny also supports successive readouts as audio arrives. The training framework uses heterogeneous supervised fine-tuning with primitive-specific objectives. Evaluation of Qwen-Audio-3.1-Jev across 32 dataset--task configurations yields 82.38\% accuracy on MMSU, 94.20\% accuracy on VoxCeleb1-O, and a NISQA overall MAE of 0.4439 MOS points. Qwen-Audio-3.1-Jev-Tiny also demonstrates complete-clip decisions and streaming support, with strengths and limitations varying across the three decision primitives. Future work will focus on two directions: supporting decisions over longer audio and expanding streaming capabilities.

\noindent\begin{minipage}{\linewidth}
\section{Authors (in Alphabetical Order of Last Name)}
\label{sec:authors}

\begin{multicols}{3}
    \begin{itemize}[noitemsep]
\item Daren Chen
\item Guoguo Chen
\item Peixin Chen
\item Xie Chen
\item Yanni Chen
\item Zhigao Chen
\item Zhiyun Fan
\item Ying Fang
\item Zhifu Gao
\item Qing Gu
\item Jiajun He
\item Yunjie Ji
\item Kangqi Jing
\item Jingbei Li
\item Xiangang Li
\item Yalin Li
\item Zengxi Li
\item Baiji Liu
\item Yaming Liu
\item Ying Liu
\item Zhaoci Liu
\item Yexin Lu
\item Yiping Peng
\item Yuezhang Peng
\item Zhendong Peng
\item Yu Pu
\item Yang Shi
\item Xin Shu
\item Xingchen Song
\item Biao Tian
\item Xiaoyu Tian
\item Peiyao Wang
\item Tianzi Wang
\item Wupeng Wang
\item Yuzhong Wu
\item Yunchong Xiao
\item Menglong Xu
\item Nan Yang
\item Jianwei Yu
\item Jixing Yu
\item Renjie Yu
\item Shimin Zhang
\item Sitong Zhao
\item Ruichen Zheng
\item Guangdong Zhou
    \end{itemize}
\end{multicols}
\end{minipage}

\begingroup
\setlength{\bibsep}{0.5pt}

\endgroup
\end{document}